\documentstyle[11pt,epsfig]{article}
\begin{document}

\title{{\bf Generalized uncertainty principle in Bianchi
type I quantum cosmology}}
\author{B. Vakili\thanks{%
email: b-vakili@sbu.ac.ir} and H. R. Sepangi\thanks{%
email: hr-sepangi@sbu.ac.ir} \\
{\small Department of Physics, Shahid Beheshti University, Evin,
Tehran 19839, Iran}} \maketitle

\begin{abstract}
We study a quantum Bianchi type I model in which the dynamical
variables of the corresponding minisuperspace obey the generalized
Heisenberg algebra. Such a generalized uncertainty principle has its
origin in the existence of a minimal length suggested by quantum
gravity and sting theory. We present approximate analytical
solutions to the corresponding Wheeler-DeWitt equation in the limit
where the scale factor of the universe is small and compare the
results with the standard commutative and noncommutative quantum
cosmology. Similarities and differences of these solutions are also
discussed. \vspace{5mm}\newline PACS numbers: 04.60.-m, 04.60.Ds,
04.60.Kz
\end{abstract}

\section{Introduction}
A general predictions of any quantum theory of gravity is that
there exists a minimal length below which no other length can be
observed \cite{1}-\cite{6}. From perturbative string theory point
of view {\cite{1,2}}, such a minimal observable length is due to
the fact that strings cannot probe distances smaller than the
sting size. In the scale of this minimal size, the quantum effects
of gravitation become as important as the electroweak and strong
interactions. Thus, in the study of high energy physics phenomena
such as the very early universe or in the strong gravitational
fields of a black hole, one cannot neglect the effects of the
existence of such a minimal length.

An important feature of the existence of a minimal length is the
modification of the standard Heisenberg commutation relation in
the usual quantum mechanics {\cite{7,8}}. Such relations are known
as the Generalized Uncertainty Principle (GUP). In one dimension,
the simplest form of such relations can be written as
\begin{equation}
\label{A}\bigtriangleup p \bigtriangleup x\geq
\frac{\hbar}{2}\left(1+\beta (\bigtriangleup
p)^2+\gamma\right),
\end{equation}
where $\beta$ and $\gamma$ are positive and independent of
$\bigtriangleup x$ and $\bigtriangleup p$, but may in general
depend on the expectation values $\langle x\rangle$ and $\langle
p\rangle$. The usual Heisenberg commutation relation can be
recovered in the limit $\beta=\gamma=0$. As is clear from equation
(\ref{A}), this equation implies a minimum position uncertainty of
$(\bigtriangleup x)_{min}=\hbar \sqrt{\beta}$, and hence $\beta$
must be related to the Planck length. For a more general
discussion on such deformed Heisenberg algebras, especially in
three dimensions, see \cite{9}. Now, it is possible to realize
equation (\ref{A}) from the following commutation relation between
position and momentum operators
\begin{equation}
\label{B} \left[x,p\right]=i\hbar \left(1+\beta
p^2\right),
\end{equation}
where we take $\gamma=\beta \langle p\rangle^2$. More general
cases of such commutation relations are studied in
\cite{7}-\cite{11}. Also various applications of the low energy
effects of the modified Heisenberg uncertainty relations have been
extensively studied, see for example \cite{12}-\cite{15}.

In this letter we consider a two dimensional minisuperspace of
Bianchi type I cosmology in the GUP framework. We shall see that
the corresponding Wheeler-DeWitt (WD) equation is a fourth order
differential equation. Although, in general we cannot solve this
equation exactly, we may obtain approximate analytical solutions
in the limit of small scale factors {\it i.e.}, in the region of
the validity of GUP. We then compare the resulting wave functions
with ordinary quantum cosmology and with noncommutative quantum
cosmology where the latter is discussed in \cite{16}.
\section{The model}
Let us consider a cosmological model in which the spacetime is
assumed to be of Bianchi type I whose metric can be written,
working in units where $c=\hbar=16\pi G=1$,  as
\begin{equation}\label{C}
ds^2=-N^2(t)dt^2+e^{2u(t)}e^{2\beta_{ij}(t)}dx^idx^j,
\end{equation}
where $N(t)$ is the lapse function, $e^{u(t)}$ is the scale factor
of the universe and $\beta_{ij}(t)$ determine the anisotropic
parameters $v(t)$ and $w(t)$ as follows
\begin{equation}\label{D}
\beta_{ij}=\mbox{diag}\left(v+\sqrt{3}w,v-\sqrt{3}w,-2v\right).
\end{equation}
To simplify the model we take $w=0$, which is equivalent to a
universe with two scale factors in the form
\begin{equation}\label{E}
ds^2=-N^2(t)dt^2+a^2(t)(dx^2+dy^2)+c^2(t)dz^2.
\end{equation}
The anisotropy in the above metric can achieved by introducing a
large scale homogeneous magnetic field in a flat FRW spacetime.
Such a magnetic field results in a preferred direction in space
along the direction of the field. If we introduce a magnetic field
which has only a $z$ component, the resulting metric can be
written in the form (\ref{E}) where there are equal scale factors
in the transverse directions $x$ and $y$ and a different one,
$c(t)$, in the longitudinal direction $z$. The dynamics of such a
universe is considered in \cite{EJ}.

Now, using the Einstein-Hilbert action
\begin{equation}\label{F}
{\cal S}=\int d^4x\sqrt{-g}({\cal R}-\Lambda),
\end{equation}
where $g$, ${\cal R}$ and $\Lambda$ represent the determinant of
the metric tensor, the scalar curvature and the cosmological
constant respectively, we are led to the following Lagrangian in
the minisuperspace $\{u,v\}$
\begin{equation}\label{G}
{\cal
L}=\frac{6e^{3u}}{N}\left(-\dot{u}^2+\dot{v}^2\right)-\Lambda N
e^{3u}.
\end{equation}
Thus, with the choice of the harmonic time
gauge $N=e^{3u}$ \cite{17}, the Hamiltonian can be written as
\begin{equation}\label{H}
{\cal H}=\frac{1}{24}\left(-p_u^2+p_v^2\right)+\Lambda
e^{6u}.
\end{equation}

Let us now proceed to quantize the model. In ordinary canonical
quantum cosmology, use of the usual commutation relations
$[x_i,p_j]=i\delta_{ij}$, results in the well known representation
$p_i=-i\partial/\partial x_i$, from which the WD equation can be
constructed. However, in the GUP framework, as was mentioned in
the introduction, the existence of a minimal observable length
requires new commutation relations between position and momentum
operators. In more than one dimension a natural generalization of
equation (\ref{B}) is defined by the following commutation
relations {\cite{12,13,18}}
\begin{equation}\label{I}
\left[x_i,p_j\right]=i\left(\delta_{ij}+\beta
\delta_{ij}p^2+\beta'p_ip_j\right),
\end{equation}
where $p^2=\sum p_ip_i$ and $\beta,\beta'>0$ are considered as
small quantities of first order. Also, assuming that
\begin{equation}\label{J}
\left[p_i,p_j\right]=0,
\end{equation}
the commutation relations for the coordinates are obtained as
\begin{equation}\label{L}
\left[x_i,x_j\right]=i\frac{(2\beta-\beta')+(2\beta+\beta')\beta
p^2}{1+\beta p^2}\left(p_ix_j-p_jx_i\right).
\end{equation}
As it is clear from the above expression, the coordinates do not
commute. This means that to construct the Hilbert space
representations, one cannot work in position space. It is
therefore more convenient to work in momentum space, as is done in
\cite{12}, \cite{13} and \cite{18}. However, since in quantum
cosmology the wave function of the universe in momentum space has
no suitable interpretation, we restrict ourselves to the special
case $\beta'=2\beta$. As one can see immediately from equation
(\ref{L}), the coordinates commute to first order in $\beta$ and
thus a coordinate representation can be defined. Now, it is easy
to show that the following representation of the momentum operator
in position space satisfies relations (\ref{I}) and (\ref{J})
(with $\beta'=2\beta$) to first order in $\beta$
\begin{equation}\label{M}
p_i=-i\left(1-\frac{\beta}{3}\frac{\partial^2}{\partial
x_i^2}\right)\frac{\partial}{\partial x_i}.
\end{equation}
A comment on the above issue is that applying the GUP to a curved
background such as a cosmological model needs some modifications
\cite{19}. Here, since we apply the GUP to the minisuperspace
variables $u,v$ which correspond to a Minkowskian metric, we can
safely use the above expressions without any modifications.
\section{Quantization of the model in the GUP framework}
Let us focus attention on the study of the quantum cosmology of
the model described by the Hamiltonian (\ref{H}). The
corresponding commutation relations are as follows
\begin{equation}\label{M}
\left[u,p_u\right]=i\left(1+\beta p^2+2\beta
p_u^2\right),\hspace{.5cm}\left[v,p_v\right]=i\left(1+\beta
p^2+2\beta p_v^2\right),\end{equation}
\begin{equation}\label{N}
\left[u,p_v\right]=\left[v,p_u\right]=2i\beta p_u
p_v,\end{equation}
\begin{equation}\label{O}
\left[x_i,x_j\right]=\left[p_i,p_j\right]=0,\hspace{.5cm}x_i(i=1,2)=u,v,
\hspace{.5cm}p_i(i=1,2)=p_u,p_v.
\end{equation}
As we have mentioned in the previous section, in the special case
when $\beta'=2\beta$, we have the following representations for
$p_u$ and $p_v$ in the $u-v$ space which satisfy the commutation
relations (\ref{M})-(\ref{O})
\begin{equation}\label{P}
p_u=-i\left(1-\frac{\beta}{3}\frac{\partial^2}{\partial
u^2}\right)\frac{\partial}{\partial
u},\hspace{.5cm}p_v=-i\left(1-\frac{\beta}{3}\frac{\partial^2}{\partial
v^2}\right)\frac{\partial}{\partial v}.
\end{equation}
Now, using these representations for momenta in the Hamiltonian
constraint (\ref{H}), the WD equation can be written, up to first
order in $\beta$, as
\begin{equation}\label{R}
\left\{\frac{\partial^2}{\partial u^2}-\frac{2}{3}\beta
\frac{\partial^4}{\partial u^4}-\frac{\partial^2}{\partial
v^2}+\frac{2}{3}\beta \frac{\partial^4}{\partial v^4}+24\Lambda
e^{6u}\right\}\Psi(u,v)=0.
\end{equation}
In the case of $\beta=0$, the ordinary quantum cosmology of the
model can be recovered and its eigenfunctions can be written in
terms of Bessel functions as follows \cite{16}
\begin{equation}\label{S}
\Psi_{\nu}(u,v)=e^{-3\nu
v}J_{\nu}\left(2\sqrt{\frac{2\Lambda}{3}}e^{3u}\right),\hspace{.5cm}\Lambda
>0,\end{equation}
\begin{equation}\label{T}
\Psi_{\nu}(u,v)=e^{3i\nu
v}K_{i\nu}\left(2\sqrt{\frac{2|\Lambda|}{3}}e^{3u}\right),\hspace{.5cm}\Lambda
<0.
\end{equation}
The solutions of equation (\ref{R}) are separable and may be
written in the form $\Psi(u,v)=U(u)V(v)$, leading to
\begin{equation}\label{U}
\frac{2}{3}\beta \frac{d^4 V}{dv^4}-\frac{d^2V}{dv^2}+9\nu^2
V=0,\end{equation}
\begin{equation}\label{V}
-\frac{2}{3}\beta \frac{d^4 U}{du^4}+\frac{d^2
U}{du^2}+\left(24\Lambda
e^{6u}-9\nu^2\right)U=0,
\end{equation}
where for having well-defined functions we use the positive
separation constant $9\nu^2$ in the case of a positive
cosmological constant. Equation (\ref{U}) is a fourth order linear
differential equation whose solutions can be written in the form
of exponential functions $e^{rv}$ where $r$ is the root of
equation
\[
2\beta r^4-3r^2+27\nu^2=0,\] where
\[r^2=\frac{3}{4\beta}\left[1\pm \left(1-24\beta
\nu^2\right)^{1/2}\right].\] To achieve the correct limit for
$\beta \rightarrow 0$, we take the negative sign in the above
expression. Thus, up to first order in $\beta$, the solution of
equation (\ref{U}) reads
\begin{equation}\label{W}
V(v)=e^{-3\nu (1+3\beta \nu^2)v}.
\end{equation}
Now, let us deal with equation (\ref{V}). Taking $\beta=0$ in this
equation yields the ordinary WD equation where its solutions in
the case of a positive $\Lambda$ are given by equation (\ref{S}).
In the case when $\beta\neq 0$, we cannot solve equation (\ref{V})
exactly, but we can provide an approximate method whose domain of
validity is given by the solution of a second order differential
equation. To this end, we note that the effects of $\beta$ become
important at the Planck scale, or in cosmology language in the
very early universe, that is, when the scale factor is small,
$e^{3u}\rightarrow 0$. Thus, if we substitute  solution (\ref{S})
in the $\beta$-term of equation (\ref{V}), we may obtain
approximate analytical solutions in the region $e^{3u}\rightarrow
0$. The limiting behavior of solution (\ref{S}) in the region
$e^{3u} \rightarrow 0$ is \cite{20}
\begin{equation}\label{Z}
J_{\nu}\left(2\sqrt{\frac{2\Lambda}{3}}e^{3u}\right)\rightarrow
\frac{1}{\Gamma(\nu+1)}\left(\sqrt{\frac{2\Lambda}{3}}\right)^{\nu}e^{3\nu
u},
\end{equation}
and therefore its fourth derivative is
\[\frac{d^4U}{du^4}=(3\nu)^4 U.\] Substituting this result into
equation (\ref{V}) leads to the following equation
\begin{equation}\label{AB}
\frac{d^2U}{du^2}+\left[24\Lambda e^{6u}-9(\nu^2+6\beta
\nu^4)\right]U=0,
\end{equation}
with solution, up to first order in $\beta$, as
\begin{equation}\label{AC}
U(u)=J_{\nu (1+3\beta
\nu^2)}\left(2\sqrt{\frac{2\Lambda}{3}}e^{3u}\right).
\end{equation}
For a negative cosmological constant, the above procedure leads to
the solutions
\[e^{3i\nu (1-3\beta \nu^2)v},\] for $V(v)$ and modified Bessel
function \[K_{i\nu (1-3\beta
\nu^2)}\left(2\sqrt{\frac{2\Lambda}{3}}e^{3u}\right),\] for
$U(u)$. Thus, the eigenfunctions of the WD equation can be written
as
\begin{equation}\label{AD}
\Psi_{\nu}(u,v)=e^{-3\nu (1+3\beta \nu^2)v}J_{\nu (1+3\beta
\nu^2)}\left(2\sqrt{\frac{2\Lambda}{3}}e^{3u}\right),\hspace{.5cm}\Lambda
>0,\end{equation}
\begin{equation}\label{AE}
\Psi_{\nu}(u,v)=e^{3i\nu (1-3\beta \nu^2)v}K_{i\nu (1-3\beta
\nu^2)}\left(2\sqrt{\frac{2|\Lambda|}{3}}e^{3u}\right),\hspace{.5cm}\Lambda
<0.\end{equation}We may now write the general solutions to the WD
equation as a superposition of the eigenfunctions
\begin{equation}\label{AF}
\Psi(u,v)=\int_{-\infty}^{+\infty} C(\nu)\Psi_{\nu}(u,v)d
\nu,\end{equation}where $C(\nu)$ can be chosen as a shifted
Gaussian weight function $e^{-a(\nu-b)^2}$.
\section{Comparison of the results}
\begin{figure}
\begin{tabular}{ccc} \epsfig{figure=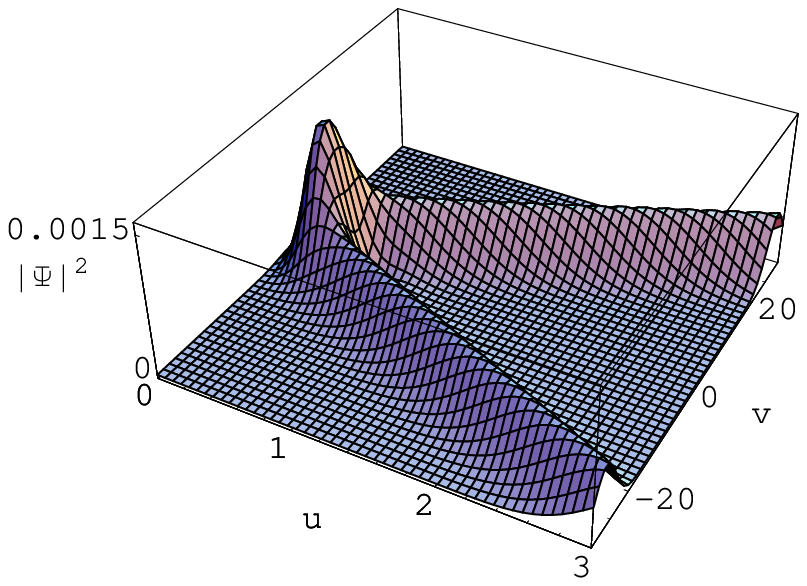,width=7cm}
\hspace{1cm} \epsfig{figure=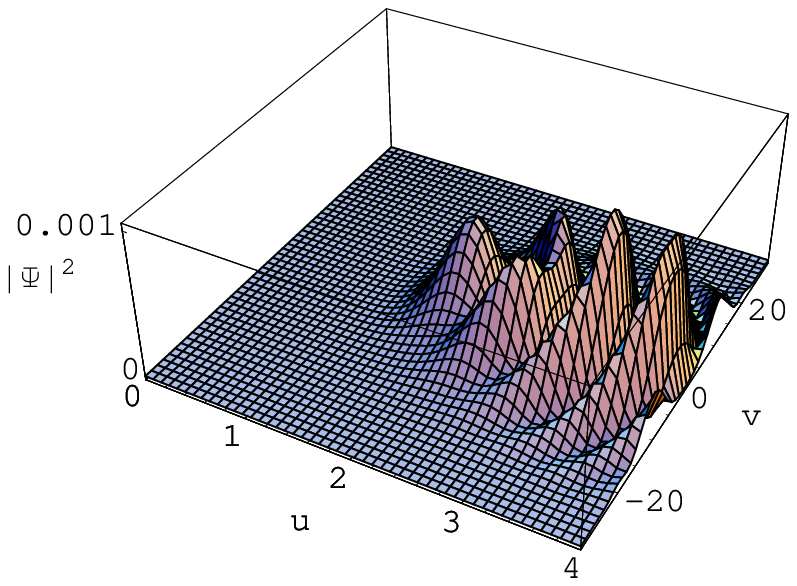,width=7cm}
\end{tabular}
\caption{\footnotesize The figure on the left shows the square of
the commutative wave function while the figure on the right, the
square of the noncommutative wave function. The figures are
plotted for a negative cosmological constant.} \label{fig1}
\end{figure}
In a previous work \cite{16}, we obtained the solutions of the
same problem in a noncommutative phase space with the following
commutation relations
\begin{equation}\label{AG}
\left[u_{nc},v_{nc}\right]=i\theta,\hspace{.5cm}\left[u_{nc},p_u\right]=
\left[v_{nc},p_v\right]=i,\hspace{.5cm}\left[p_u,p_v\right]=0,
\end{equation}
and showed that the eigenfunctions of the corresponding WD
equation in such a space are given by
\begin{equation}\label{AH}
\Psi_{\nu}(u,v)=e^{-3\nu
v}J_{\nu}\left(2\sqrt{\frac{2\Lambda}{3}}e^{3(u-\frac{3}{2}\nu
\theta)}\right),\hspace{.5cm}\Lambda >0,
\end{equation}
\begin{equation}\label{AI}
\Psi_{\nu}(u,v)=e^{3i\nu v}K_{i\nu}\left(2\sqrt{\frac{2|\Lambda
|}{3}}e^{3(u-\frac{3}{2}\nu \theta)}\right),\hspace{.5cm} \Lambda
<0.
\end{equation}
As can be seen, the general solutions are again in the form of an
expression like (\ref{AF}). Although in the present study and with
the choice $\beta'=2\beta$, the phase space variables $u, v$
commute with each other, in general equation (\ref{L}) shows that
GUP leads naturally to a noncommutative generalization of position
space which may point to a close relationship between
noncommutativity and GUP. The study of such phenomena as IR/UV
mixing and non-locality, Lorentz violation and new physics at very
short distances in  noncommutative \cite{21}-\cite{25} and GUP
{\cite{26,27}} frameworks, would pave the way for a more clear
understanding of this relationship.

\begin{figure}\begin{center}
\epsfig{figure=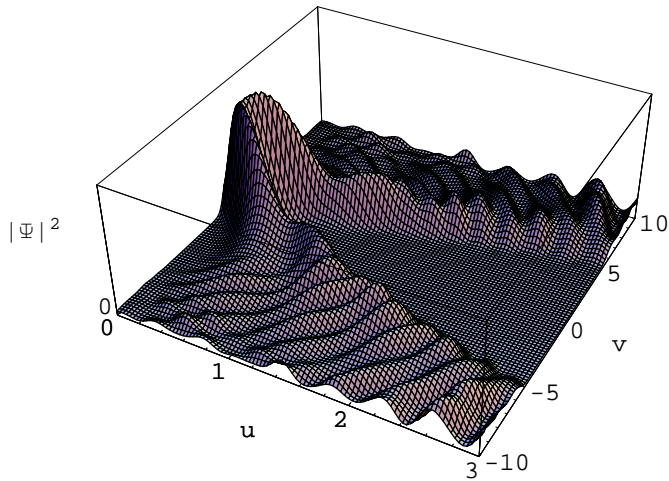,width=9cm} \caption{\footnotesize The
square of wave function in the GUP framework. The figure is
plotted for a negative cosmological constant.}
\label{fig2}\end{center}
\end{figure}

Figures \ref{fig1} and \ref{fig2} show the square of wave
functions in the context of an ordinary commutative phase space,
noncommutative phase space and when the phase space variables obey
the GUP relations. As is clear from these figures, in the ordinary
commutative case ($\theta=\beta=0$), we have only one possible
universe around a nonzero value of $u$ and $v=0$, which means that
the universe in this case approaches a flat FRW one. On the other
hand we see that noncommutativity causes a shift in the minimum
value of $u$ corresponding to the spatial volume. The emergence of
new peaks in the noncommutative wave packet may be interpreted as
a representation of different quantum states that may communicate
with each other through tunnelling. This means that there are
different possible universes (states) from which our present
universe could have evolved and tunnelled, from one state to
another (see also \cite{28}). Such behavior also occurs in figure
\ref{fig2} which shows the square of the GUP wave function,
showing that from the point of view adopted here, noncommutativity
and GUP may be considered as similar concepts. However, there is
an important difference, namely, that the noncommutative wave
function not only peaks around $v=0$, but appear symmetrically
around a nonzero value of $v$, which is the characteristic of an
anisotropic universe. On the other hand, the GUP wave function as
is seen in the figure, has many peaks around the value $v=0$ and
therefore from the point of predicting an isotropic universe the
GUP wave packet behaves like the ordinary commutative case.
\section{Conclusions}
In this paper we have studied the effects of the existence of a
minimal length scale on the quantum states of a Bianchi type I
cosmology. This phenomena yields a deformed Heisenberg commutation
relation between the position and momentum operators which are
known as the GUP. Although in more than one dimension, because of
the existence of a noncommutative relation between the space
operators it is not possible to represent the momentum operator in
the position space, in the special case when the GUP parameters
obey the relation $\beta'=2\beta$, the space operators commute and
a coordinate representation of momenta can be defined. Since a
wave function describing the quantum state of the universe does
not not have a suitable interpretation in momentum space, working
in the above special case is crucial in our work.

When the phase space variables obey the ordinary commutation
relations we have seen that there is only one possible isotropic
universe. Upon considering the GUP and through finding the
approximate analytical solutions of the WD equation in the limit
of small scale factors, we have studied the corresponding quantum
cosmology and seen that in the presence of the GUP, the square of
the wave function of the universe has several peaks. This
behavior, also occurring  in the noncommutative quantum model
studied in a previous work \cite{16}, may be related to different
states in the early universe from which our present universe could
have evolved. Although, both noncommutativity and GUP predict many
possible initial universes, this is not the case in predicting the
isotropicity. In the case of a noncommutative cosmology the
universe behaves anisotropicaly  while the GUP predicts an
isotropic universe.

\end{document}